\documentclass[conference]{IEEEtran}
\IEEEoverridecommandlockouts
\usepackage{lmodern}
\usepackage{booktabs}
\usepackage{moreverb}
\usepackage{fontenc}
\usepackage{amsmath}
\usepackage{amssymb}
\usepackage{fancybox}
\usepackage{wrapfig}
\usepackage{color}
\usepackage{colortbl}
\usepackage{array}
\usepackage{cite}
\usepackage{natbib}
\usepackage{multirow}
\usepackage{multicol}
\usepackage{listings}
\usepackage{makecell}
\usepackage{graphicx}
\usepackage{rotating}
\usepackage{setspace}
\usepackage{soul}

\usepackage{balance}
\usepackage{csvsimple}
\usepackage{longtable}
\usepackage[skip=0pt,labelfont=bf]{caption}
\usepackage{url}
\usepackage{lscape}
\usepackage{rotating}
\usepackage{tikz}
\usepackage{enumitem}
 \usepackage{subfigure}
\usepackage{wrapfig}
\usepackage[most]{tcolorbox}

\usepackage{threeparttable}

\usepackage{bclogo}
\usepackage{multicol}
\usepackage{marvosym}
\usepackage{pifont}

\newcolumntype{L}[1]{>{\raggedright\let\newline\\\arraybackslash\hspace{0pt}}m{#1}}
\newcolumntype{C}[1]{>{\centering\let\newline\\\arraybackslash\hspace{0pt}}m{#1}}
\newcolumntype{R}[1]{>{\raggedleft\let\newline\\\arraybackslash\hspace{0pt}}m{#1}}

\definecolor{codegreen}{rgb}{0,0.6,0}
\definecolor{codegray}{rgb}{0.5,0.5,0.5}
\definecolor{codepurple}{rgb}{0.58,0,0.82}
\definecolor{backcolour}{rgb}{0.95,0.95,0.92}
\definecolor{amber}{rgb}{0.91, 0.84, 0.42}

\lstdefinestyle{mystyle}{
    commentstyle=\color{codegreen},
    keywordstyle=\color{magenta},
    numberstyle=\tiny\color{codegray},
    stringstyle=\color{codepurple},
    basicstyle=\footnotesize,
    breakatwhitespace=false,
    breaklines=true,
    captionpos=b,
    keepspaces=true,
    showspaces=false,
    showstringspaces=false,
    showtabs=false,
    tabsize=2
}

\setlist{noitemsep} 

\definecolor{darkpastelred}{rgb}{0.76, 0.23, 0.13}
\definecolor{ao(english)}{rgb}{0.0, 0.5, 0.0}

\definecolor{darkpastelred}{rgb}{0.76, 0.23, 0.13}
\definecolor{ao(english)}{rgb}{0.0, 0.5, 0.0}

\lstdefinelanguage{diff}{
  morecomment=[f][\color{blue}]{@@},     
  morecomment=[f][\color{red}]-,         
  morecomment=[f][\color{codegreen}]+,       
  morecomment=[f][\color{red}]{---}, 
  morecomment=[f][\color{codegreen}]{+++},
}

\definecolor{yellow}{RGB}{255,255,153}
\definecolor{grey}{RGB}{224,224,224}

\newboolean{showcomments}
\setboolean{showcomments}{true}
\ifthenelse{\boolean{showcomments}}
 { \newcommand{\mynote}[2]{
      \fbox{\bfseries\sffamily\scriptsize#1}
        {\small$\blacktriangleright$\textsf{\emph{#2}}$\blacktriangleleft$}}}
        { \newcommand{\mynote}[2]{}}

\definecolor{DarkOrange}{rgb}{0.8,0.3,0.0}
\definecolor{DarkCyan}{rgb}{0.0, 0.55, 0.55}

\newcolumntype{?}{!{\vrule width 1pt}}

\def\BibTeX{{\rm B\kern-.05em{\sc i\kern-.025em b}\kern-.08em
    T\kern-.1667em\lower.7ex\hbox{E}\kern-.125emX}}
\begin{document}

\title{\huge Clean Code In Practice: Challenges and Opportunities}


\author{\footnotesize
  \IEEEauthorblockN{%
  Dapeng Yan\IEEEauthorrefmark{1}, 
    Wenjie Yang\IEEEauthorrefmark{2},
    Kui Liu\IEEEauthorrefmark{3}, 
    Zhiming Liu\IEEEauthorrefmark{4},
    Zhikuang Cai\IEEEauthorrefmark{1}, 
  }
  \vspace{0.5em} 
  \IEEEauthorblockA{\footnotesize\IEEEauthorrefmark{1}%
    College of Integrated Circuit Science and Engineering,
    Nanjing University of Posts and Telecommunications, Nanjing, China\\
    \{dapeng.yan, whczk\}@njupt.edu.cn%
  }
  \vspace{0.3em}
  \IEEEauthorblockA{\footnotesize\IEEEauthorrefmark{2}%
    School of Computer Science and Technology,
    China University of Mining and Technology, Xuzhou, China\\
    TS23170054A31@cumt.edu.cn%
  }
  \vspace{0.3em}
  \IEEEauthorblockA{\footnotesize\IEEEauthorrefmark{3}%
    Huawei Technologies Co., Ltd., Hangzhou, China \\
    brucekuiliu@gmail.com%
  }
  \vspace{0.3em}
  \IEEEauthorblockA{\footnotesize\IEEEauthorrefmark{4}%
    School of Software Engineering, Southwest University, Chongqing, China \\
    zhimingliu88@swu.edu.cn%
  }
}

\maketitle

\vspace{-600em}  

\begin{abstract}
Reliability prediction is crucial for ensuring the safety and security of software systems, especially in the context of industry practices. While various metrics and measurements are employed to assess software reliability, the complexity of modern systems necessitates a deeper understanding of how these metrics interact with security and safety concerns. This paper explores the interplay between software reliability, safety, and security, offering a comprehensive analysis of key metrics and measurement techniques used in the industry for reliability prediction. We identify critical threats to software reliability and provide a threat estimation framework that incorporates both safety and security aspects. Our findings suggest that integrating reliability metrics with safety and security considerations can enhance the robustness of software systems. Furthermore, we propose a set of actionable guidelines for practitioners to improve their reliability prediction models while simultaneously addressing the security and safety challenges of contemporary software applications.
\end{abstract}


\section{Introduction}
\label{sec: intro}
``{\emph{Even bad code can function. However, if code is not clean, it can bring a development organization to its knees}}''~\cite{martin2009clean}. 
Clean code is generally accepted as code that can be easily understood by developers other than its original author(s). Understandability in this context encompasses readability, testability, changeability, extensibility, and code maintainability. Therefore, writing clean code has become a crucial skill for every developer to master~\cite{ccexplain}.

A substantial body of work in the literature discusses the concept of clean code and the conventions for writing it.
Robert C. Martin believes that writing clean code requires a disciplined use of various small techniques applied through a painstakingly acquired sense of "cleanliness," and he thinks that the key is "code sense."
He presents a revolutionary paradigm in the book ``Clean Code: A Handbook of Agile Software Craftsmanship''~\cite{martin2009clean} with the understanding of the best agile practice of clean code from 16 aspects (e.g., meaningful names, functions, and comments), which is popularly recommended in the literature and industry.
In this book, Robert C. Martin reports the understanding of clean code from some well-known programming language experts whom he interviewed before writing the book.  
This includes the opinion of Bjarne Stroustrup that "elegant code" is about readability and efficiency. The idea of Dave Thomas is that clean code is associated with {\emph{readability}}, {\emph{tests}} (code without tests is not clean), and {\emph{sizes}} (smaller is better).

Industry practitioners also present their understanding and practice of clean code. For example, technical experts from Google considered that clean code should possess seven characteristics: consistent, non-duplicative, efficient and scalable, simple and direct, maintainable, optimized for the reader, and leave an explicit trace for the reader~\cite{google}.
Microsoft researchers consider that the intrinsic characteristics of clean code should include comprehensibility, correctness, consistency, advancement, safety, and security~\cite {microsoft}.

In this paper, we present our empirical study on the state of clean code in the industry and the challenges faced by developers.
To the best of our knowledge, this is the first systematic study that explores the current state and issues surrounding clean code.
We begin by investigating the extent to which industry code is clean, analyzing code from Apache, Google, and Microsoft across four aspects: function sizes, file size, code line length, and naming conventions. These analyses aim to provide reference metrics for clean code practices in the community.
Then, we distribute questionnaires to gather developers' opinions and suggestions on implementing clean code in practice. 
The main contributions of this work include:
\begin{itemize}[leftmargin=*]
    \item The referable and measurable metrics, with four aspects (i.e., function size, file size, code line lengths, and naming convention), for clean code can be extracted from industry code, providing potential baselines for clean code practice.
    \item After analyzing responses from online questionnaires, we find that developers' complaints mainly focus on the one-size-fits-all mandatory policy of practicing clean code and the incorrect role of automated tools played in clean code practice.
\end{itemize}

\section{Empirical Study}
\label{sec: sd}

This empirical study is conducted with two aspects to answer the following research questions:
\begin{itemize}[leftmargin=*]
	\item \textbf{RQ-1.} {\bf \emph{Is it possible to carry out the potential referable baselines from the industry to evaluate clean code for developers?}}
	It would be better to have referable baselines to provide potential guidelines for clean code development.
	To that end, we investigate the current state of clean code from four perspectives: function sizes, file sizes, code line lengths, and naming conventions. 
	\item \textbf{RQ-2.} {\bf \emph{What kinds of challenges are faced by developers in the process of proceeding with clean code?}}
	For this RQ, we propose inquiring into the developers' opinions on clean code who are engaged in developing clean code in the industry with questionnaires.
\end{itemize}

\subsection{Study Design}

To answer the research questions, we conducted two separate studies, the related designs of which are presented below.

\subsubsection{\bf Subject Selection for RQ-1}
For RQ-1, we consider selecting the representative data.
In the literature, open-source code from Apache, Google, and Microsoft has been widely studied as a subject for various tasks~\cite{sadowski2018modern,devanbu2016belief}. 
Additionally, Google and Microsoft make an effort to adhere to clean code practices.
Therefore, we select the top 10 most popular (highest number of stars) open-source projects for each of the five widely used programming languages (i.e., C/C++, Java, JavaScript, and Python) from GitHub, as maintained by Apache, Google, and Microsoft, respectively. 
The details of our analyzed data are shown in Table~\ref{tab:dataset_source_only}.

\begin{table}[!h]
  \centering
  \small
  \vspace{-2mm}
  \caption{The statistics of functions and lines of selected subjects (only source files).}
  \resizebox{\linewidth}{!}{
    \begin{tabular}{l|rr|rr|rr}
    \toprule
    \multirow{2}{*}{Language}
      & \multicolumn{2}{c|}{Apache}
      & \multicolumn{2}{c|}{Google}
      & \multicolumn{2}{c}{Microsoft} \\
    \cline{2-7}
      & \# Functions & LOCs
      & \# Functions & LOCs
      & \# Functions & LOCs \\
    \hline
    C     &      152,044  &   4,603,941
          &       13,971  &     377,860
          &      646,260  &  16,827,882 \\
    C++   &       91,033  &   1,656,318
          &       26,635  &     587,748
          &       43,884  &     740,669 \\
    Java  &      342,549  &   6,061,173
          &      169,785  &   1,538,751
          &       60,108  &     808,659 \\
    JavaScript
          &        6,424  &     276,320
          &       36,508  &     964,648
          &       10,049  &   1,297,325 \\
    Python
          &       42,841  &     893,470
          &       44,987  &     621,461
          &       14,838  &     305,680 \\
    \hline
    Total &      634,891  &  13,491,222
          &      291,886  &   4,090,468
          &      775,139  &  19,980,215 \\
    \bottomrule
    \end{tabular}%
  }
  \label{tab:dataset_source_only}
  \vspace{-3mm}
\end{table}

\subsubsection{\bf Questionnaires for RQ-2}
\label{sec:Design-rq-2}
For RQ-2, we deliver online questionnaires to collect opinions from programmers on clean code, and the related methods are illustrated below.

\textbf{\emph{Protocol.}}
Our questionnaires are delivered on Credamo, which is a one-stop smart research online platform that can support the accurate push of questionnaires about the following two questions:
\begin{enumerate}
    \item {\bf What kinds of problems on clean code practice are faced by developers?}
    \item {\bf Are there any suggestions from developers for the practice of clean code?} (It is an optional question.)
\end{enumerate}

\textbf{\emph{Participant Selection.}} In this study, we select the participants for questionnaires with four criteria: \ding{172} who have participated in answering questionnaires at least 50 times, \ding{173} whose credit score (i.e., the metric used in Credamo to measure the quality of user answers) is higher than or equal to 80, \ding{174} whose adoption rate of historical answers is not lower than 70\%, and \ding{175} who has at least 10-year experience on software development.

\textbf{\emph{Valid Responses.}}
We pushed our questionnaires to 700 developers through the platform.
At the end of April 20, 2025, we received 537 responses.
Invalid answers, such as ``\emph{ I do not know}'', were excluded.
Eventually, 460 valid results with a valid rate of 85.7\% are collected.

{\bf\emph{Categories of Responses.}}
As outlined in the ``Protocol,'' our questionnaire addresses two aspects: problems and suggestions. For the problems, we use the \textit{Open Card Sorting} method to categorize valid responses, which follows three steps:
\ding{172} \textbf{Response Analysis}: We first organize the valid responses into a table, listing each response separately. Three authors independently analyze the responses and categorize them based on thematic similarities, with themes emerging from the responses themselves, rather than predefined categories~\cite{latoza2006maintaining}.
\ding{173} \textbf{Agreement Examination}: To assess the consistency among annotators, we calculate the Fleiss Kappa score~\cite{egele2013empirical}. The Kappa value of 0.77 indicates substantial agreement among the annotators.
\ding{174} \textbf{Disagreement Discussion}: In case of discrepancies, the annotators discuss and reach a consensus on the classification.
To minimize bias in theme sorting, we reviewed and agreed on a final set of 5 main categories and 18 subcategories (cf. Section~\ref{sec:rq2}). This classification method is also applied to the suggestions from the second aspect of the questionnaire. The annotators then finalize the classifications through agreement.

\section{Study Results}
\label{sec:sr}

\begin{figure}[!h]
    \vspace{-5mm}
    \includegraphics[width=\linewidth]{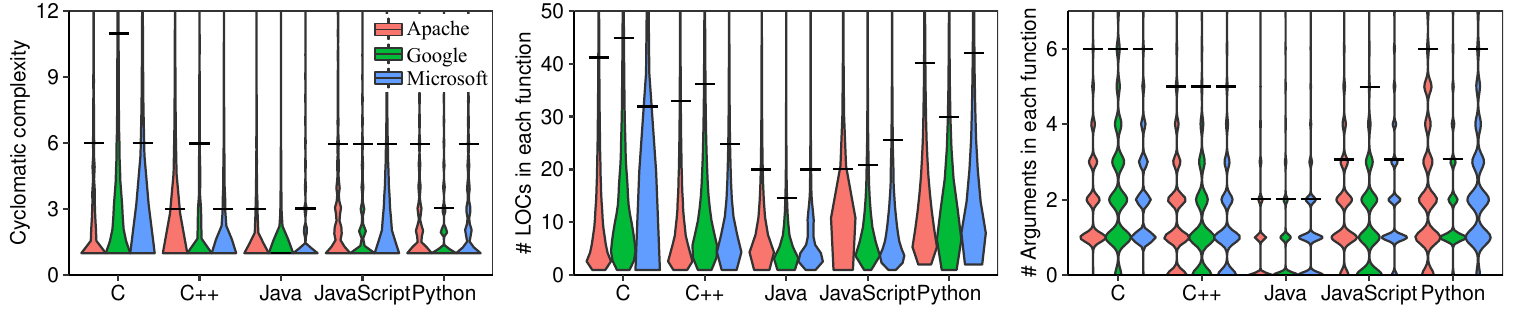}
    \caption{{Distribution of function sizes.}}
    \vspace{-5mm}
    \label{fig:func}
\end{figure}

In this section, we investigate clean code practices by analyzing function sizes, file sizes, code line lengths, and naming conventions in 150 open-source projects from Apache, Google, and Microsoft.
We then discuss the challenges that developers face in clean code practice.

\subsection{RQ-1: Measurable Baselines of Clean Code}
\label{sec:rq1}
In this study, we investigate the current situation of program code developed by Apache, Google, and Microsoft from four aspects: function size, file size, line length, and naming convention, to provide measurable indicators for clean code development.
Function sizes include cyclomatic complexity~\cite{mccabe1976cyclomatic} and lines of code (LOCs) in each function, while file sizes are presented with lines of code in each source file.
The code line length presents the number of code tokens in each line.
Naming conventions are evaluated by checking whether variable, function, and class names adhere to ``{\bf CamelCase}'' and ``{\bf snake\_case}'' formats.

Uncle Bob mentioned that the function should be small, and the name should be meaningful when writing clean code~\cite{martin2009clean}.
As shown in Table~\ref{tab:my_label}, Google recommends limiting code line length to 80 characters for C, C++, JavaScript, and Python, and 100 characters for Java code lines.
In McCabe's presentation 'Software Quality Metrics to Identify Risk'~\cite{McCabe}, he interpreted cyclomatic complexity with four categories.
If the cyclomatic complexity of a function belongs to the range of 1 to 10, it indicates a simple procedure with little risk.

\begin{table}[!t]
    \centering
    \vspace{1mm}
    \caption{Recommended code line lengths, LOCs and cyclomatic complexity of functions.}
    \label{tab:my_label}
    \resizebox{0.9\linewidth}{!}{
    \begin{threeparttable}
    \begin{tabular}{c|c|c|c|c|c}
        \toprule
        Google & C & C++ & Java & JavaScript & Python \\\hline
        code line lengths & 80 & 80 & 100 & 80 & 80 \\
        LOCs of functions & 40 & - & - & - & 40\\\hline\hline
        \multicolumn{6}{c}{Cyclomatic Complexity}\\\hline
        1 - 10   & \multicolumn{5}{c}{Simple procedure, little risk}\\\hline
        11 - 20 & \multicolumn{5}{c}{More complex, moderate risk}\\\hline
        21 - 50 & \multicolumn{5}{c}{Complex, high risk}\\\hline
        $>$ 50    & \multicolumn{5}{c}{Untestable code, very high risk}\\
        \bottomrule
    \end{tabular}
    \vspace{-9mm}
    \end{threeparttable}
    }
\end{table}

\subsubsection{Function Sizes}
\label{fs}

Figure~\ref{fig:func} illustrates the distribution of function sizes in terms of the cyclomatic complexity, LOCs of functions, and the number of function arguments.
Overall, most functions satisfy the small function criterion of clean code, of which cyclomatic complexities are limited within 5, LOCs of functions are within 40, and the number of function arguments is limited to 6.
The median and mean values of related metrics are much lower, [1, 2] and [2, 3] for the cyclomatic complexities, [3, 10] and [6, 19] for LOCs of functions, and [0, 2] and [1, 3] for number of function arguments, for each language, respectively.

Function sizes of one language present similar distributions.
For the C language functions from Apache, Google, and Microsoft, their cyclomatic complexities are mainly distributed in the intervals of [1, 6], [1, 11], and [1, 6], respectively.
Moreover, their LOCs are mainly distributed in the intervals of [1, 41], [1, 45], and [1, 32].
The number of function arguments is distributed in the same range of [0, 6].
Similar value intervals for related metrics indicate that functions written in the same language exhibit similar distributions in their sizes, despite being from different companies. 

To sum up, most program functions are developed within a small range, which is consistent with the small-function principle of clean code that is practiced by the three companies themselves and recommended by professional experts~\cite{martin2009clean}.
Their corresponding values can provide measurable and reproducible baselines. However, such baselines should be measured with concrete references for different languages.

\subsubsection{File Sizes}

\begin{wrapfigure}{r}{0.2\textwidth}
  \vspace{-2mm}  
  \centering
  \includegraphics[width=\linewidth]{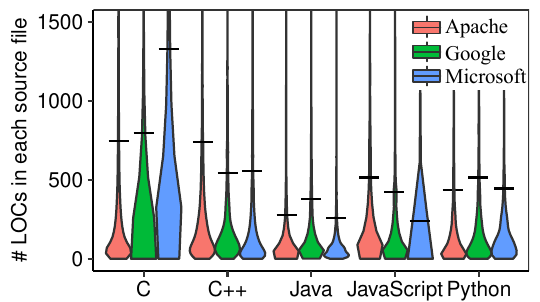}
  \caption{File sizes}
  \label{fig: FS}
  \vspace{-4mm}
\end{wrapfigure}

Figure~\ref{fig: FS} shows the distribution of file sizes in terms of LOCs of source code files.
The three companies with five programming languages normally encapsulate all the source code within 600 code lines (the third quartile).
Examining the box sizes, the source code files of the C language present a significantly wider range of file sizes than those of the other four languages.
It might be a consequence that C is a procedure-oriented programming language.
Overall, file sizes present a similar distribution to function sizes.

With these statistics, we can also conclude that most code files are encapsulated within a small range of code lines, which is consistent with the concise code criterion of clean code practice.
Their related values can provide measurable and referable baselines for participants who are working on developing code files for clean code practices. However, such baselines should also be appropriately varied for the specific programming language.

\subsubsection{Code Line Lengths}
\begin{wrapfigure}{r}{0.2\textwidth}
  \vspace{-3mm}
  \centering
  \includegraphics[width=\linewidth]{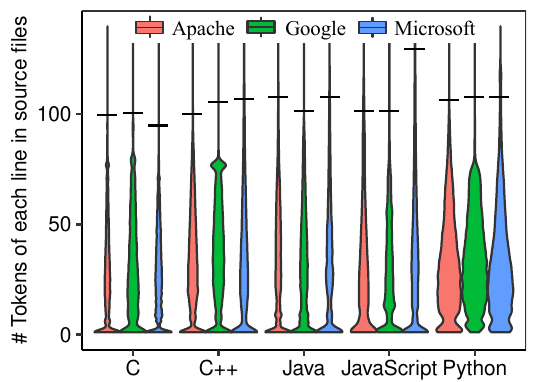}
  \caption{Code line lengths}
  \label{fig: LOCs}
  \vspace{-4mm}
\end{wrapfigure}
Overall, as shown in Figure~\ref{fig: LOCs}, most code lines of source code are normally coded with 1 (e.g., ``{\tt \{}'') to $\sim$100 characters.
This suggests that developers tend to write concise code, with each line containing a small number of characters.

In conclusion, most code line lengths are developed within a narrow range, which is consistent with the clean code conventions that the three companies adhere to.
Their related values can provide measurable and referable baselines for practicing clean code at the granularity of a code line.
\subsubsection{Outlier Distributions}
According to the distributions mentioned above, we note that the sizes of some code lines, functions, and files are out of the related normal size ranges. We further look into these outliers.
Except for setting thresholds based on the previous distributions, we also refer to the corresponding sizes recommended by companies for their clean code implementations.
The statistics of outliers that are out of the corresponding sizes (i.e., thresholds, the upper whisker values of the related data) are presented in Table~\ref{tab: thresh}.

\begin{table*}[!h]
  \centering
  \caption{Thresholds and the ratio of outliers for cyclomatic complexities, LOCs, number of arguments of functions and files, and code line lengths.}
    \resizebox{\linewidth}{!}{
    \begin{threeparttable}
    \begin{tabular}{l|rrr|rrr|rrr|rrr|rrr}
    \hline
    {Language} & \multicolumn{3}{c|}{C} & \multicolumn{3}{c|}{C++} & \multicolumn{3}{c|}{Java} & \multicolumn{3}{c|}{JavaScript} & \multicolumn{3}{c}{Python} \\
    \hline
    {Company} & {Apache} & {Google} & {Microsoft} & {Apache} & {Google} & {Microsoft} &{Apache} & {Google} &{Microsoft} & {Apache} & {Google} & {Microsoft} & Apache & {Google} & {Microsoft} \\
    \hline
    {Cyclomatic complexity} & 6     & 11    & 6     & 3   & 6     & 3     & 3     & 1     & 3     & 6     & 6     & 6     & 6     & 3     & 6 \\
    Ratio of outliers (\%) & 12.10 & 9.25 & 10.44 & 15.85 & 9.87 & 13.09 & 10.06 & 17.00 & 11.10 & 11.32 & 7.66 & 8.85 & 8.89 & 7.96 & 11.02 \\
    \hline
    {LOCs of functions } & 41    & 45    & 32    & 33    & 36 (40)    & 24    & 20    & 14    & 20    & 20    & 21    & 25    & 40     & 30     & 42 \\
    Ratio of outliers (\%) & 8.99 & 9.03 & 8.52 & 8.70& 8.35(6.93) & 10.06& 9.33& 7.27& 6.93& 7.92& 8.75& 7.40& 7.67& 7.01& 7.46\\
    \hline
    {{\# Function Arguments}} & 6     & 6     & 6     & 5     & 5     & 5     & 2     & 2     & 2     & 3.5   & 5     & 3     & 6     & 3     & 6 \\
    {Ratio of outliers} (\%) & 2.01& 3.58& 1.41& 2.44& 1.77& 2.44& 8.21& 5.35& 7.13& 8.70& 0.67& 11.85& 5.43& 11.31& 4.92\\
    \hline
    {LOCs of source code files} & 781   & 831   & 1332  & 775   & 546   & 561   & 280   & 388   & 232   & 519   & 425   & 206   & 434   & 527   & 439 \\
    Ratio of outliers (\%) & 11.26& 7.03& 8.90& 8.57& 11.12& 10.42& 9.55& 10.72& 9.18& 9.91& 8.03& 10.26& 10.21& 9.38& 6.74\\
    \hline
    Ratio of outliers (\%) & 11.46& 12.23& 11.23& 9.37& 11.59& 11.79& -  & -  & -  & -  & -  & -  & -  & -  & - \\
    \hline
    {Code line lengths} & 98    & 105 (100)   & 88    & 102   & 111 (80)   & 116   & 117   & 106 (100)   & 117   & 106   & 106 (80)   & 129   & 93    & 96 (80)    & 97 \\
    Ratio of outliers (\%)  & 0.61& 0.22(0.28) & 0.35& 0.15& 0.04(0.55) & 1.65& 0.53& 0.03(0.05) & 0.65& 1.52& 2.52(4.74) & 4.28& 0.89& 0.20(0.66) & 2.42\\
    \hline
    Ratio of outliers (\%)  & 0.18& 0.02& 0.06& 0.98& 0.71& 2.65& -  & -  & -  & -  & -  & -  & -  & -  & - \\
    \hline
    \end{tabular}%
    {$^\ast$(\#) - numbers in the parentheses are the sizes (i.e., thresholds) recommended by the related company for clean code implementation.}
    \end{threeparttable}
    }
  \label{tab: thresh}
  \vspace{-7mm}
\end{table*}%

We observe that the upper whisker value (36) is close to the function size (40) recommended by Google for its Java clean code.
The upper whisker values of other cases are also close to this recommended function size or even are much lower than it.
Google also lists the recommended code line lengths, and the related values are lower than the related upper whisker values. However, ratios of outliers do not change so high (cf. the ``Code line lengths'' row in Table~\ref{tab: thresh}) when comparing the two different thresholds.
These results further suggest that most developers are willing to practice clean code in their development, adhering to concise criteria.

The outliers of code line lengths present much lower ratios than the function sizes and file sizes.
For example, the outliers of code lines in code files developed by Google using the C language occupy only 0.22\% and 0.02\% ratios, respectively, which are significantly lower than the ratios of outliers (9.25\%, 9.03\%, 7.03\%, and 12.23\%) for the corresponding function and file sizes.
Code lines are much simpler than functions and files, so it is easier to write clean code lines than clean code functions and files.
It thus raises a challenge for clean code practice.

Google provides six baselines (i.e., thresholds presented in parentheses of Table~\ref{tab: thresh}) for LOCs of C++ functions and code line lengths of the five languages.
Their related outliers present lower ratios than related functions and code line lengths from Apache and Microsoft.
When baselines are not provided for cyclomatic complexity, file sizes, and function sizes of C, C++, JavaScript, and Python, the outlier ratios from Google are typically higher than those of the other two companies. 
For example, the LOCs of C functions in Google present a higher outlier ratio than those in Apache and Microsoft.
It indicates that clear baselines can guide developers to better code practices.

Apache, Google, and Microsoft have achieved benchmark-level clean code with concise criteria.
Their developers are prone to write clean code, and their code could be cleaner if measurable baselines are provided.
Nevertheless, large functions, files, and lengthy code lines still exist in their code repositories, which can affect the readability and maintainability of programs.
Refactoring huge code fragments into concise code is also a challenge for them.

\subsubsection{Naming Conventions}
In this study, we further investigate the implementation status of naming conventions among these companies.
To this end, we first construct pattern-matching regular expressions based on the grammar of each language to extract class names, function names, and variable names.
Since C and JavaScript have no concept of classes, we consider analyzing the naming conventions of their filenames. 
In the end, we collected a total of 180,555 class names, 1,663,515 function names and 3,044,319 variable names for this investigation.
Then, we use \textbf{``CamelCase''} and {\bf ``snake\_case''} (i.e., under\_score) naming formats to check whether these names satisfy these two conventions that are widely recommended in the community~\cite{liu2019learning}.
If a name cannot be matched to both formats, it is identified as an irregular name.

Table~\ref{tab:cname} illustrates the distributions of irregular class, function, and variable names,
All three companies face the problem of irregular names.
The ratios of irregular class names in Java and JavaScript code developed by Microsoft are much higher than in other cases.
For function names, Apache and Google share the same issue with the high number of irregular function names in C and C++ code. In contrast, Microsoft presents a similar issue for C++ and JavaScript code.
For irregular variable names, Java code in Microsoft presents a much higher ratio than others.
Nevertheless, there are a large number of irregular variable names in the Apache code of Java, JavaScript, and C, as well as in the Google code of Java, Python, and JavaScript, and the Microsoft code of Python and C.

\begin{table}[!h]
  \centering
  \vspace{-2mm}
  \caption{Distributions of irregular class, function, and variable names.}
  \resizebox{0.85\linewidth}{!}{
    \begin{tabular}{lc|ccccc}
    \toprule
    & & C & C++ & Java & JavaScript & Python \\\hline
    \multirow{3}{*}{\makecell[l]{irregular\\class\\names}} & \multicolumn{1}{|c|}{Apache}
    & 0.21\% & 2.96\% & 0.07\% & 0 & 2.59\% \\
    \cline{2-7}
         &  \multicolumn{1}{|c|}{Google} & 0.19\% &  0  & 0.68\%  &  3.04\%  & 3.41\% \\
    \cline{2-7}
         & \multicolumn{1}{|c|}{Microsoft} & 0.4\% & 1.11\% & 17.73\% & 9.75\% & 1.96\%  \\
    \hline
    \multirow{3}{*}{\makecell[l]{irregular\\function\\names}} & \multicolumn{1}{|c|}{Apache} 
    & 15.9\% & 14.92\% & 0.1\% & 0.22\% & 3.53\% \\
    \cline{2-7}
         & \multicolumn{1}{|c|}{Google} & 15.57\% & 13.25\% & 0.52\% & 6.81\% & 4.72\% \\
    \cline{2-7}
         & \multicolumn{1}{|c|}{Microsoft} & 5.23\% & 16.78\% & 5.74\% & 21.76\% & 0.49\%  \\
    \hline
    \multirow{3}{*}{\makecell[l]{irregular\\variable\\names}} & \multicolumn{1}{|c|}{Apache} 
    & 0.61\% & 2.56\% & 2.76\% & 1.25\% & 2.64\% \\
    \cline{2-7}
         & \multicolumn{1}{|c|}{Google} & 1.53\% & 0.74\% & 2.51\% & 1.15\% & 3\% \\
    \cline{2-7}
         & \multicolumn{1}{|c|}{Microsoft} & 0.29\% & 2.15\% & 9.85\% & 0.28\% & 3.19\%  \\
    \hline
    \end{tabular}%
    }
  \label{tab:cname}%
    \vspace{-4mm}
\end{table}%

\subsection{RQ-2: Challenges and Opportunities Arisen From Developers' Clean Code Practices}
\label{sec:rq2}

We follow the method presented in Section~\ref{sec:Design-rq-2} to categorize the responses into main categories and their corresponding subcategories. The detailed classifications of the main categories are described below: 
\begin{itemize}[leftmargin=*]
    \item {\bf Management} category denotes the responses complaining about the company's proceeding plan on clean code implementation and the overall management policies of practicing and validating clean code.
    
    \item {\bf Constraints} category means the various constraints of assessing to what extent the code is clean (e.g., the size of a function should be less than 50 LOCs).
    
    \item {\bf Tool} category represents complaints from respondents that are related to tools used in clean code practice.
    
    \item {\bf Committer} category refers to responses discussing clean code things concerning code change commits and merge requests.
    
    \item {\bf Test} category represents the test discussed in responses.
\end{itemize}

\subsubsection{Complaints From Developers}
\label{sec:pp}

\begin{figure}[!ht]
  \centering
  \vspace{-2mm}
  \begin{minipage}[t]{0.48\linewidth}
    \centering
    \vspace{0pt} 
    \includegraphics[width=\linewidth]{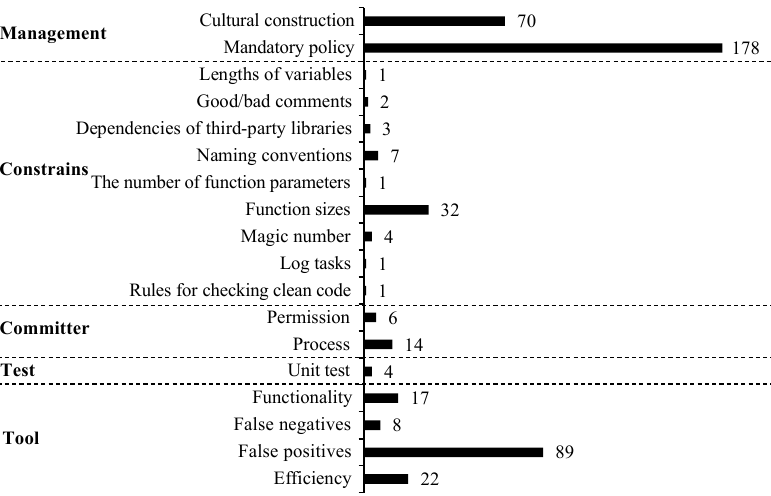}
    \captionof{figure}{Distribution of sub-categories.}
    \label{fig:subtypes}
  \end{minipage}
  \hfill
  \begin{minipage}[t]{0.44\linewidth}
    \centering
    \vspace{0pt} 
    \captionof{table}{Distribution of main categories.}
    \resizebox{\linewidth}{!}{
      \begin{tabular}{lcc}
        \toprule
        Main Type & Count & Proportion \\
        \midrule
        Management & 248 & 53.91\% \\
        Tool       & 136 & 29.57\% \\
        Convention &  52 & 11.30\% \\
        Committer  &  20 &  4.35\% \\
        Test       &   4 &  0.87\% \\
        \midrule
        Total      & \multicolumn{2}{c}{460} \\
        \bottomrule
      \end{tabular}%
    }
    \label{tab:vrc}
  \end{minipage}
  \vspace{-8mm}
\end{figure}

Table~\ref{tab:vrc} presents the distribution of the five main categories.
Most responses are concentrated in the two categories of management and tool, which account for 84\% (=54\% + 30\%) of all valid responses.
It indicates that most developers are concerned about the overall actionable execution strategy and the tools that assist developers.

Figure 4 reveals the distribution of sub-categories of developers' complaints. 
For the {\bf Management} category, it includes 70 and 178 responses on cultural construction (the cultivation and atmosphere of learning and developing clean code for its implementation) and mandatory policy (the overall policy of implementing and validating clean code), respectively.
The mandatory policy presents the highest number of developers' complaints, as many are unhappy with the one-size-fits-all and strict policies for validating clean code.
For example, their companies set mandatory baselines on the function size of 50 code lines, which is an impossible task for them when they work on specific, difficult programming tasks.
Additionally, writing clean code is directly tied to developers' key performance indicators (KPI), causing them to feel huge pressure and to regard clean code as a forced task driven by performance metrics rather than as an inspiring art.

In the Constraints category, developers complain most about the strict 50-line function rule; despite its simplicity, it is the most challenging to implement in practice. They ask why the threshold cannot be higher or adjusted for different programming languages. Three potential reasons underlie this antipathy:  \ding{172} mandatory constraints without guidance, so that even a one‑ or two‑line overrun triggers penalties; \ding{173} the enormous effort required to simplify already‑optimized functions; and \ding{174} the delay in delivery caused by extensive refactoring.

Developers complain about the naming conventions of clean code: \ding{172} it is hard to come up with different function names with a limited number of characters to distinguish similar functions in large projects.
\ding{173} It is inevitable to invoke APIs with ``unclean'' names from the third-party libraries because these packages could be legacy programs or developed by other teams or companies. However, the clean code checking tools cannot distinguish the invoked APIs and developer-coined names, which leads to inaccurate assessments of unclean code.

{\bf Committer} contains two sub-categories: permission and process, which refer to the power (who can act) and workflow (what is the complete procedure) of reviewing commits, respectively. 
In some software product development teams, when their software products are being delivered, reviewing code change commits is a highly controlled process.
Respondents complain that committers do not have sufficient permission to review commits, which prevents developers from promptly assessing the quality of changed code.
Developers also complain that the commit reviewing process even requires approval from the chairman of the software product (the waiting time for such approval is always long), which complicates the tasks that can be handled in a simple and fast way.

Testability is also a major concern. Developers struggle to unit‑test legacy code because essential information is often missing, especially in older codebases. Mandatory coverage targets then force them to write tests for the sake of coverage rather than quality or efficiency, resulting in redundant, overlapping tests that make the code “unclean.” This challenge underscores the necessity for automated test-generation tools for legacy systems and methods to detect and eliminate test overlap, thereby improving both test quality and efficiency.

We categorize the complaints about {\bf Tool} into four sub-categories: functionality, efficiency, false positives, and false negatives related to other tools.
From the responses to the questionnaires, we note that some developers leverage code-checking tools to help them practice clean code by identifying potential issues in the code.
However, such tools often report a high number of false positives, and developers thus complain that they must expend considerable effort in resolving these false positives.
Even worse, changing these false positives leads to more ``unclean'' codes.
Some developers consider that the tool was initially developed to help developers write clean code. However, their companies use the tool to assess whether their code is clean or not.
Overall, these findings reveal an urgent need for more adaptive, context‑aware clean‑code policies and tools that guide rather than mandate.

\subsubsection{Suggestions From Developers}
\begin{figure}[htbp]  
  \vspace{-2mm}                         
  \centering
  \includegraphics[width=\linewidth]{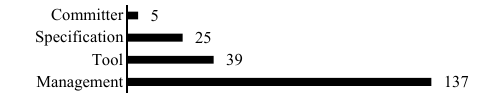}
  \caption{Distribution of suggestions in questionnaires.}
  \label{fig:suggestion}
  \vspace{-2mm}
\end{figure}

Although suggestions are optional for respondents, there are still 44\%~($\approx$204/460) of them providing their suggestions on clean code development.
The corresponding distribution is shown in Figure~\ref{fig:suggestion}.  

For the {\bf Management} category, the related suggestions focus on improving the one-size-fits-all mandatory policy.
Developers suggest that the clean code assessment management should build a professional team to create appropriate and actionable indicators for various clean code practice scenarios.
They state that companies initially take much time to establish clean code conventions and benefit greatly later. They recommend that such conventions should be involved in the cultural construction of clean code as much as possible.

For the {\bf Tool} category, the suggestions mainly discuss what role the tool should play in clean code development.
Developers suggest that tools are used to assist programmers with clean code practice, but not to assess programmers' code.
Developers also confirm that they indeed benefit from the automated tools for clean code and suggest that the tools should be upgraded and optimized according to the issues that occurred in clean code practice.
The other suggestions related to the {\bf Constraints} and {\bf Committer} are mainly about the flexible and guiding baselines considering the clean code implementation, but not the one-size-fits-all mandatory metrics, and the actionable and convenient commit reviewing procedures.

In conclusion, these suggestions from developers also imply that developers are willing to practice clean code in their programming development tasks. Moreover, they genuinely contribute their efforts to boost clean code development.
The challenges faced by developers arise in various domains for enhancing clean code, including establishing actionable, referable, and measurable baselines for clean code practices, as well as automated techniques to assist programmers in developing clean code.

\section{Related Work}
\label{sec:rw}

\textbf{Code Quality.} Clean code is a well-defined method and standard for code quality~\cite{rachow2018missing}. Developers and users have analyzed the quality of code from different aspects. Tonella and Abebe define availability and performance as external quality~\cite{tonella2008code} because they are dynamic and generally measured at runtime. Instead, internal quality, such as maintainability, is more likely to reflect in the static structure of the software~\cite{tonella2008code}~\cite{mari2003impact}. MARI and EILA, another descriptive term group, defined evolution as the internal quality and execution as the external quality~\cite{mari2003impact}.
Stamelos et al.~\cite{stamelos2002code} presented a result from an experimental case study that aims to understand the meaning of structural quality and clarify the results of structural quality analysis of code delivered to open-source development. Moreover, Allamanis et al.~\cite{allamanis2014learning} proposed ``Naturalize'', a framework for learning the code library and recommending modification suggestions to improve coding styles.

\textbf{Code Convention.} Martin provides numerous conventions in his book~\cite{martin2009clean} for writing clean code. However, Li and Prasad believe that although developers understand the importance of using code conventions, they often fail to follow them when the development task needs to be completed quickly~\cite{li2005effectively}. SMIT et al.~\cite{smit2011maintainability} disagree and propose an indicator to analyze maintenance in different fields, namely ``agreement'' indicators based on these coding agreed violations.

Unlike these works, we are the first to investigate the challenges and opportunities of clean code in practice.
Our empirical study on analyzing open-source projects from Apache, Google, and Microsoft aims to investigate the current state of clean code practices in the industry, providing results that can serve as referable and measurable indicators for corresponding practitioners.
The analysis of our questionnaire further discusses the challenges of clean code practice faced by developers and the potential opportunities that arise from it.

\section{Conclusion}
\label{conclusion}

Clean code has garnered considerable attention from practitioners.
In this work, we conducted an empirical study on clean code to investigate its current application status from four aspects using open-source projects from Apache, Google, and Microsoft. We discussed the challenges faced by developers and the opportunities identified after analyzing our questionnaire results.
Although clean code development has encountered various challenges, there are still ways to address them, and developers are willing to adopt clean code and are ready to welcome the benefits that it brings. 

\scriptsize
\bibliographystyle{unsrt}
\bibliography{bib/references}

\begin{thebibliography}{10}

\bibitem{martin2009clean}
Robert~C Martin.
\newblock {\em Clean code: a handbook of agile software craftsmanship}.
\newblock Pearson Education, 2009.

\bibitem{ccexplain}
Yi{\u g}it~Kemal Erin{\c c}.
\newblock Clean code explained – a practical introduction to clean coding for beginners.
\newblock \url{https://www.freecodecamp.org/news/clean-coding-for-beginners/}, Last Access: August 2022.

\bibitem{google}
Google.
\newblock 10 tipis for clean code.
\newblock \url{https://techdevguide.withgoogle.com/resources/topics/clean-code/}, Last Access: August 2022.

\bibitem{microsoft}
Microsoft.
\newblock All in one code framework.
\newblock \url{https://learn.microsoft.com/en-us/shows/onecode/}, Last Access: August 2022.

\bibitem{sadowski2018modern}
Caitlin Sadowski, Emma S{\"o}derberg, Luke Church, Michal Sipko, and Alberto Bacchelli.
\newblock Modern code review: a case study at google.
\newblock In {\em Proceedings of the 40th International Conference on Software Engineering: Software Engineering in Practice}, pages 181--190, 2018.

\bibitem{devanbu2016belief}
Premkumar Devanbu, Thomas Zimmermann, and Christian Bird.
\newblock Belief \& evidence in empirical software engineering.
\newblock In {\em 2016 IEEE/ACM 38th International Conference on Software Engineering (ICSE)}, pages 108--119. IEEE, 2016.

\bibitem{latoza2006maintaining}
Thomas~D LaToza, Gina Venolia, and Robert DeLine.
\newblock Maintaining mental models: a study of developer work habits.
\newblock In {\em Proceedings of the 28th international conference on Software engineering}, pages 492--501, 2006.

\bibitem{egele2013empirical}
Manuel Egele, David Brumley, Yanick Fratantonio, and Christopher Kruegel.
\newblock An empirical study of cryptographic misuse in android applications.
\newblock In {\em Proceedings of the 2013 ACM SIGSAC conference on Computer \& communications security}, pages 73--84, 2013.

\bibitem{mccabe1976cyclomatic}
Thomas~J McCabe~Sr.
\newblock Cyclomatic complexity.
\newblock {\em National Bureau of Standards. special Publication. m99}, 1976.

\bibitem{McCabe}
Tom McCabe.
\newblock Software quality metrics to identify risk.
\newblock \url{http://www.mccabe.com/ppt/SoftwareQualityMetricsToIdentifyRisk.ppt}, Last Access: August 2022.

\bibitem{liu2019learning}
Kui Liu, Dongsun Kim, Tegawend{\'e}~F. Bissyand{\'e}, Anil Koyuncu, Kisub Kim, Taeyoung Kim, Suntae Kim, and Yves Le~Traon.
\newblock Learning to spot and refactor inconsistent method names.
\newblock In {\em Proceedings of the 41st ACM/IEEE International Conference on Software Engineering}, pages 1--12. IEEE, 2019.

\bibitem{rachow2018missing}
Paula Rachow, Sandra Schr{\"o}der, and Matthias Riebisch.
\newblock Missing clean code acceptance and support in practice-an empirical study.
\newblock In {\em 2018 25th Australasian Software Engineering Conference (ASWEC)}, pages 131--140. IEEE, 2018.

\bibitem{tonella2008code}
Paolo Tonella and Surafel~Lemma Abebe.
\newblock Code quality from the programmer’s perspective.
\newblock In {\em Proceedings of Science, XII Advanced Computing and Analysis Techniques in Physics Research, Erice, Italy}, volume~5, page 153, 2008.

\bibitem{mari2003impact}
Matinlassi Mari et~al.
\newblock The impact of maintainability on component-based software systems.
\newblock In {\em Euromicro Conference}, pages 25--25. IEEE Computer Society, 2003.

\bibitem{stamelos2002code}
Ioannis Stamelos, Lefteris Angelis, Apostolos Oikonomou, and Georgios~L Bleris.
\newblock Code quality analysis in open source software development.
\newblock {\em Information systems journal}, 12(1):43--60, 2002.

\bibitem{allamanis2014learning}
Miltiadis Allamanis, Earl~T. Barr, Christian Bird, and Charles Sutton.
\newblock Learning natural coding conventions.
\newblock In {\em Proceedings of the 22nd {ACM} {SIGSOFT} International Symposium on Foundations of Software Engineering}, pages 281--293. {ACM}, 2014.

\bibitem{li2005effectively}
Xiaosong Li and Christine Prasad.
\newblock Effectively teaching coding standards in programming.
\newblock In {\em Proceedings of the 6th conference on Information technology education}, pages 239--244, 2005.

\bibitem{smit2011maintainability}
Michael Smit, Barry Gergel, H~James Hoover, and Eleni Stroulia.
\newblock Maintainability and source code conventions: An analysis of open source projects.
\newblock {\em University of Alberta, Department of Computing Science, Tech. Rep. TR11}, 6, 2011.

\end{thebibliography}

\end{document}